\documentclass[a4paper,11pt]{article}
\pdfoutput=1

\usepackage{jheppub}

\newcommand{\bea}{\begin{eqnarray}}
\newcommand{\eea}{\end{eqnarray}}

\makeatletter
\@addtoreset{equation}{section}
\makeatother

\title{
Phase Diagram of a Holographic Superconductor \\
Model with s-wave and d-wave  
}
\author{
Mitsuhiro Nishida[a]
}

\affiliation[a]
{Department of Physics, Graduate School of Science,
Osaka University, Toyonaka, Osaka 560-0043, Japan}

\emailAdd{nishida(at)het.phys.sci.osaka-u.ac.jp}

\abstract{We consider a holographic model with a scalar field, a tensor field and a direct coupling between them as a superconductor with an s-wave and a d-wave. We find a rich phase structure in the model. The model exhibits a phase of coexistence of the s-wave and the d-wave, or a phase of an order competition. Furthermore, it has a triple point.
}

\preprint{OU-HET 809
}

\begin{document}
\maketitle
\setcounter{page}{1}

\section{Introduction}
\label{sec1}

Recently, AdS/CFT correspondence \cite{Maldacena:1997re, Gubser:1998bc, Witten:1998qj} in superstring theory was studied actively. Its application to other physics is expected to be an innovative method. For example, it has been applied to nuclear physics and condensed matter physics (see, for example, \cite{Kovtun:2004de, McGreevy:2009xe}). Holographic superconductor is one of the applications. A motivation for the holographic superconductor is to better-understand physics from the relation between gauge theories and gravity theories. After a simple scalar model was shown to have a property characteristic to superconductors or superfluids \cite{HHH}, various models have been studied. 

One example of interesting superconductors is an anisotropic superconductor \cite{Ueda}. In condensed matter physics, a rotational symmetry on an angular momentum of a cooper pair is important. A cooper pair of some superconductors, such as a copper oxide \cite{Scalapino}, has nonzero angular momentum and they are called as anisotropic superconductor. Motivated by interest in gauge/gravity correspondence of it, holographic models of a vector field or a tensor field were studied \cite{Gubser:2008wv, Cai:2013aca, Chen:2010mk, Benini:2010pr, Kim:2013oba}.         

Other example is a multi band superconductor such as $\textrm{MgB}_2$ and iron pnictides \cite{Yamashita, Stewart}. Furthermore, there are superconductors such as $\textrm{CePt}_3\textrm{Si}$ \cite{Fujimoto} in which two order parameters whose symmetries are different from each other coexist. To describe them, holographic models with two or more fields corresponding to the order parameters were studied \cite{Basu:2010fa, Wen:2010et, Huang:2011ac, Krikun:2012yj, Musso:2013ija, Cai:2013wma, Liu:2013yaa, Wen:2013ufa, Nie:2013sda, Amado:2013lia, Amoretti:2013oia, Donos:2013woa, Nitti:2013xaa}.

In this paper, we consider a holographic model with a scalar field, a tensor field and a direct coupling between them as a superconductor with an s-wave and a d-wave. We find a rich phase structure in the model. The model exhibits a phase of coexistence of the s-wave and the d-wave, or a phase of an order competition. Furthermore, it has a triple point.

The authors of Ref.~\cite{Benini:2010pr} have shown that a d-wave model reduces to a scalar model under a specific ansatz. Therefore, our result can be applied also for a two-scalar model. The two-scalar model was studied first in Ref.~\cite{Basu:2010fa} and it was expected that the regime of the coexistence phase changes depending on the values of a direct coupling. We specifically confirm it by numerical calculations.

The organization of this paper is as follows. In section 2, we explain our holographic model with the s-wave and the d-wave and its equations of motion. In section 3, we calculate solutions of the model and their free energy densities, and study their properties. In section 4, we study the properties of the phases for a range of values of the direct coupling and see that our model has the rich phase structure as shown in figure 7. Section 5 is for a summary and discussions.
%%%%%%%%%%%%%%%%%%%%%%%%%%%%%%%%%%%%%%%%%%%%%%%%%%%%%

\section{Our gravity model}
\label{sec2}
In this section, we explain our holographic superconductor model with an s-wave and a d-wave. Some holographic models with a scalar field or a tensor field are studied previously \cite{HHH, Chen:2010mk, Benini:2010pr, Kim:2013oba}.  These fields are interpreted as order parameters. To study coexistence and an order competition of the two order parameters, we consider a scalar field, a tensor field and a direct coupling between them.     

For exotic superconductivity, temperature is essential and two-dimensional space is considered to be indispensable. To accommodate them in holography, we usually use a four-dimensional AdS planar black hole metric
\begin{align}
ds^2=\frac{L^2}{z^2}(-f(z)&dt^2+dx^2+dy^2+\frac{dz^2}{f(z)}),\\
f(z)&=1-\Big(\frac{z}{z_h}\Big)^3,
\end{align}
where $z=0$ is the AdS boundary and $z=z_h$ is the black hole horizon. Temperature of a superconductor corresponds to the Hawking temperature $T$ of this black hole 
\begin{align}
T=\frac{3}{4\pi z_h}.
\end{align}

Lagrangians with a scalar field or a tensor field were proposed as an s-wave or a d-wave superconductor \cite{HHH, Chen:2010mk, Benini:2010pr, Kim:2013oba}. To combine them, we consider a Lagrangian with a Maxwell field $A_\mu$, a scalar field $\psi$, a symmetric tensor field $\Phi_{\mu\nu}$ and a direct coupling constant $\eta$ between $\psi$ and $\Phi_{\mu\nu}$  as
\begin{align}
S=&\int d^4x\sqrt{-g}\left[-\frac{1}{4}F^{\mu\nu}F_{\mu\nu}+\mathcal{L}_\textrm{s}+\mathcal{L}_\textrm{d}-\eta|\psi|^2|\Phi_{\mu\nu}|^2\right],\\
\mathcal{L}_\textrm{s}=&-m_\textrm{s}^2|\psi|^2-|D_\mu\psi|^2,\\
\mathcal{L}_\textrm{d}=&-|D_\rho\Phi_{\mu\nu}|^2+2|D_\mu\Phi^{\mu\nu}|^2+|D_\mu\Phi|^2-[(D_\mu\Phi^{\mu\nu})^*D_\nu\Phi+\textrm{c.c.}]\notag\\
&-m_\textrm{d}^2(|\Phi_{\mu\nu}|^2-|\Phi|^2)+2R_{\mu\nu\rho\lambda}\Phi^{*\mu\rho}\Phi^{\nu\lambda}-\frac{1}{4}R|\Phi|^2-ie_\textrm{d}F_{\mu\nu}\Phi^{*\mu\lambda}\Phi^\nu_\lambda,\\
D_\mu=&\nabla_\mu-ie_aA_\mu\;\;\;\;\;(a=\textrm{s}, \textrm{d}),
\end{align}
where $\mathcal{L}_\textrm{d}$ is an effective action of a spin two field for the d-wave part \cite{Benini:2010pr}\footnote{As noted later, our model is similar to a two-scalar model and reduces to it under the ansatz (2.12). In two-scalar models, one can consider also the following Josephson coupling between two scalar fields \cite{Wen:2013ufa}
\begin{align}
\psi_1^*\psi_2+\psi_1\psi_2^* .
\end{align}
In our model, correspondingly we can consider a coupling between the scalar field and the tensor field
\begin{align}
\psi^*g^{\mu\nu}\Phi_{\mu\nu}+\psi g^{\mu\nu}\Phi_{\mu\nu}^*.
\end{align}
However, we do not consider this coupling because (2.9) is zero under the anzats (2.12). 
}. $\psi$ and $\Phi_{\mu\nu}$ correspond to the s-wave and the d-wave.

Because our main purpose is to observe variety of the phase diagram, we consider the direct coupling $\eta$ only and analyze our model in the probe limit\footnote{In this paper, we do not consider consistency of the direct coupling and the probe limit.}. This simplifies our calculations, although we lose the generality of the model.

We set the mass and the charge of the fields as
\begin{align}
&m^2_\textrm{s}L^2=-2,\;\;\;\;\;m^2_\textrm{d}L^2=0,\\
&e_\textrm{s}=1,\;\;\;\;\;\;\;\;\;\;\;\;\;\;e_\textrm{d}=1.95.
\end{align}
These parameters are same as \cite{Basu:2010fa}. We choose these parameters to compare our result with the result of \cite{Basu:2010fa}. The mass of each field corresponds to the dimension of the order parameters. $m^2_\textrm{s}$ is negative, but it does not lead to an instability because it is above the Breitenlohner-Freedman bound \cite{Breitenlohner:1982jf}. $e_\textrm{s}$ and $e_\textrm{d}$ are interpreted as effective charge of the cooper pairs. Our choice of the parameters (2.10) and (2.11) is for the solutions of our model to have a rich phase structure. In fact, $e_\textrm{s}$ and $e_\textrm{d}$ contribute to the  effective mass squared through the covariant derivative (2.7) in (2.5) and (2.6).  

For the effective action of a spin-two field, large $m^2_\textrm{d}$ is better \cite{Benini:2010pr}. However, we expect that there is not large difference about $m^2_\textrm{d}$ because our model reduces to a two-scalar model. To measure the effect of $m^2_\textrm{d}$, we will consider other values of the parameters in section 4. 

We suppose that the superconductor is homogeneous and there are isotropic and anisotropic cooper pairs. Our ansatz for the fields corresponded to them naturally is 
\begin{align}
\psi=\psi(z),\;\;\;\;\;\Phi_{xy}=\Phi_{yx}=\frac{L^2}{2z^2}\varphi(z),\;\;\;\;\;A_t=\phi(z),
\end{align}
and the other components are zero. We also set $\psi, \varphi$ and $\phi$ to be real for simplicity. Under this ansatz, our model reduces to a two-scalar model with a direct coupling only. The equations of motion are
\begin{align}
&\psi''+\Big(\frac{f'}{f}-\frac{2}{z}\Big)\psi'+\frac{e_\textrm{s}^2\phi^2}{f^2}\psi+\frac{2L^2}{z^2f}\psi-\frac{\eta L^2\varphi^2}{2z^2f}\psi=0,\\
&\varphi''+\Big(\frac{f'}{f}-\frac{2}{z}\Big)\varphi'+\frac{e_\textrm{d}^2\phi^2}{f^2}\varphi-\frac{\eta L^2\psi^2}{z^2f}\varphi=0,\\
&\phi''-\frac{2e_\textrm{s}^2L^2\psi^2}{z^2f}\phi-\frac{e_\textrm{d}^2L^2\varphi^2}{z^2f}\phi=0,
\end{align}
and asymptotic solutions of the equations of motion around the boundary $z=0$ are
\begin{align}
&\psi=\psi^{(1)}z+\psi^{(2)}z^2,\\
&\varphi=\varphi^{(1)}+\varphi^{(2)}z^3,\\
&\phi=\mu-\rho z.
\end{align}

In order to solve the equations of motion, we need boundary conditions. We set them as
\begin{align}
\psi^{(1)}=0,\;\;\;\;\;\varphi^{(1)}=0,\;\;\;\;\;\phi(z_h)=0.
\end{align} 
In holographic superconductor, coefficients of the asymptotic solutions (2.16), (2.17) and (2.18) correspond to external fields and their responses. We regard $\mu$ and $\rho$ as a chemical potential and a charge density. Moreover, we regard $\psi^{(2)}$ and $\varphi^{(2)}$ as vacuum expectation values of the order parameters
\begin{align}
\langle\mathcal{O}_\textrm{s}\rangle=\psi^{(2)},\;\;\;\;\;\langle\mathcal{O}_\textrm{d}\rangle=\varphi^{(2)}.
\end{align}
Fixing $\mu$, we will calculate behavior of $\langle\mathcal{O}_\textrm{s}\rangle$ and $\langle\mathcal{O}_\textrm{d}\rangle$ by changing $T$.

In the next section, we will find four types of solutions of the model: 
\begin{itemize}
\item solution of the normal conducting phase (the normal conducting solution)\footnote{One can check that this normal conducting solution satisfies (2.13), (2.14), (2.15) and (2.19).}, $\psi=\varphi=0$, $\phi=\mu(1-z/z_h)$.
\item solution of the s-wave superconducting phase (the s-wave single solution), $\langle\mathcal{O}_\textrm{s}\rangle\neq0, \varphi=0$.
\item solution of the d-wave superconducting phase (the d-wave single solution), $\psi=0, \langle\mathcal{O}_\textrm{d}\rangle\neq0$.
\item solution in which the s-wave superconductivity and the d-wave superconductivity coexist (the s+d coexistent solution), $\langle\mathcal{O}_\textrm{s}\rangle\neq0, \langle\mathcal{O}_\textrm{d}\rangle\neq0$.
\end{itemize}

%%%%%%%%%%%%%%%%%%%%%%%%%%%%%%%%%%%%%%%%%%%%%%%%%%%

\section{Solutions of the model}
\label{sec3}

In this section, we calculate the solutions of the equations of motion by a numerical method using Mathematica. By the symmetry of the metric (2.1), we can fix $\mu=1$. Moreover, we set $L=1$ for a numerical calculation.

%%%%%%%%%%%%%%%%
\subsection{Single solutions}
First, we calculate the single solutions. The single solutions do not depend on $\eta$, since $\psi=0$ or $\varphi=0$. A numerical result is shown in figure 1. The left blue curve is for the s-wave single solution, the right red curve is for the d-wave single solution and $T_\textrm{d}$ is the phase transition temperature at which the d-wave condensation begins. One can see that the d-wave single solution begins to condense at higher temperature than that of the s-wave single solution. Generally, if mass squared of fields $m^2$ is small, it condenses at high temperature. In our model, we set $e_\textrm{d}$ larger than $e_\textrm{s}$ and the effective mass squared of the tensor field is smaller than that of the scalar field at high temperature.

\begin{figure}[t]
\centering
\includegraphics[scale=0.7]{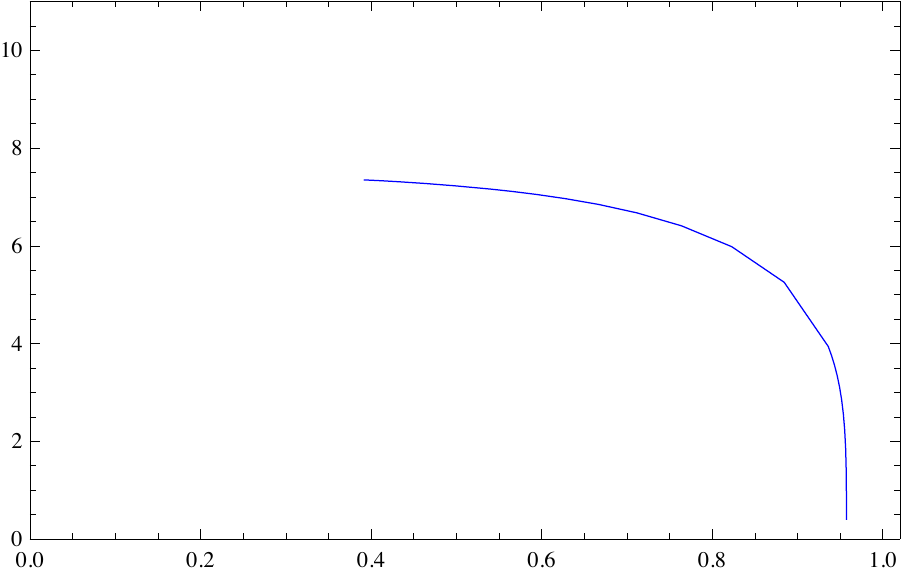} \hspace{10mm}
\includegraphics[scale=0.7]{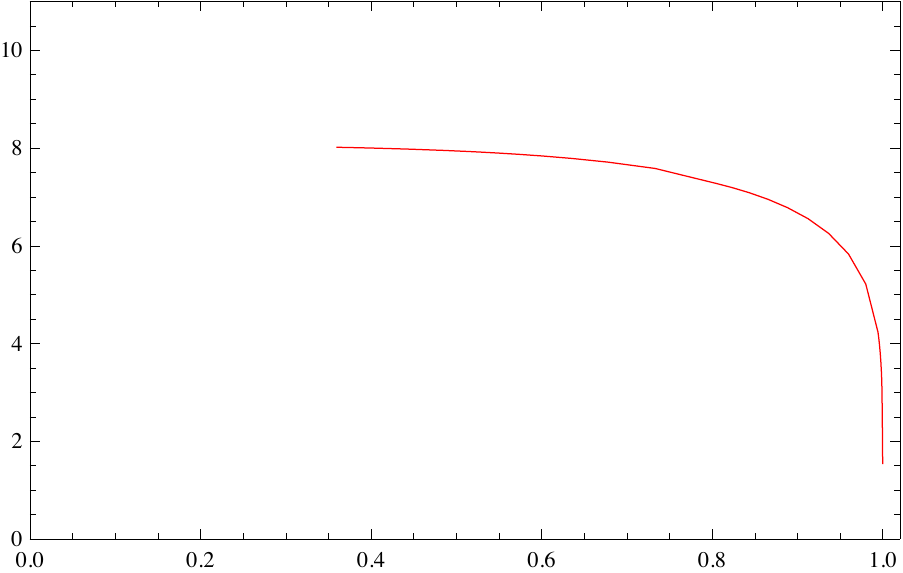}
\put(-100,-10){$T/T_\textrm{d}$}
\put(-320,-10){$T/T_\textrm{d}$}
\put(-430,60){$\frac{\langle\mathcal{O}_\textrm{s}\rangle^{1/2}}{T_\textrm{d}}$}
\put(-213,60){$\frac{\langle\mathcal{O}_\textrm{d}\rangle^{1/3}}{T_\textrm{d}}$}
\caption{Plot of the vacuum expectation values of the order parameters of the single solutions. Left figure is for the s-wave single solution and right figure is for the d-wave single solution. The axes are normalized by $T_\textrm{d}$.}
\label{fig1}
\end{figure}

%%%%%%%%%%%%%%%
\subsection{s+d coexsistent solutions}

Second, we calculate the s+d coexistent solutions. Figure 2 is for the s+d coexistent solutions of $\eta=-1/10$ and $\eta=0$. Lowering the temperature, $\langle\mathcal{O}_\textrm{s}\rangle$ becomes large and $\langle\mathcal{O}_\textrm{d}\rangle$ becomes small. The range in which the solution of $\eta=-1/10$ exists is larger than that of the solution of $\eta=0$.
\begin{figure}[t]
\centering
\includegraphics[scale=0.7]{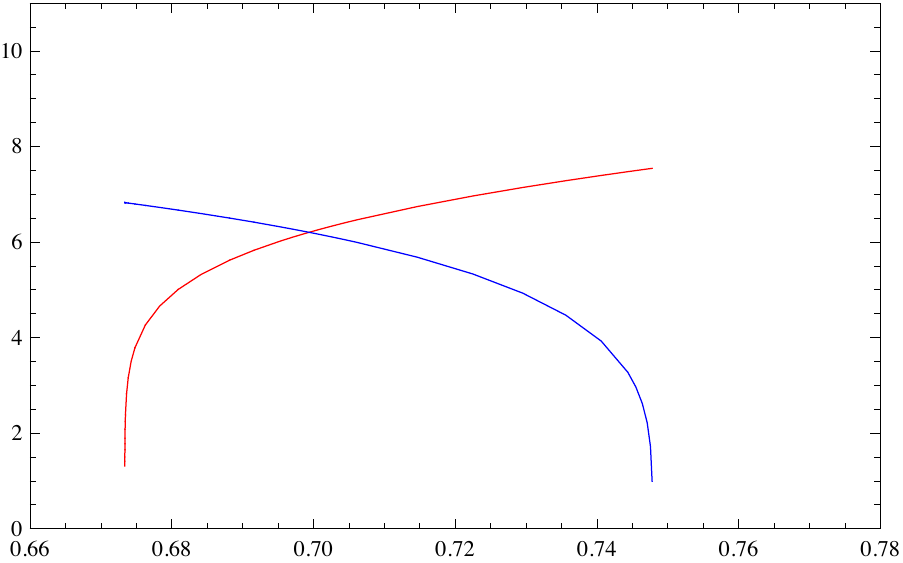} \hspace{10mm}
\includegraphics[scale=0.7]{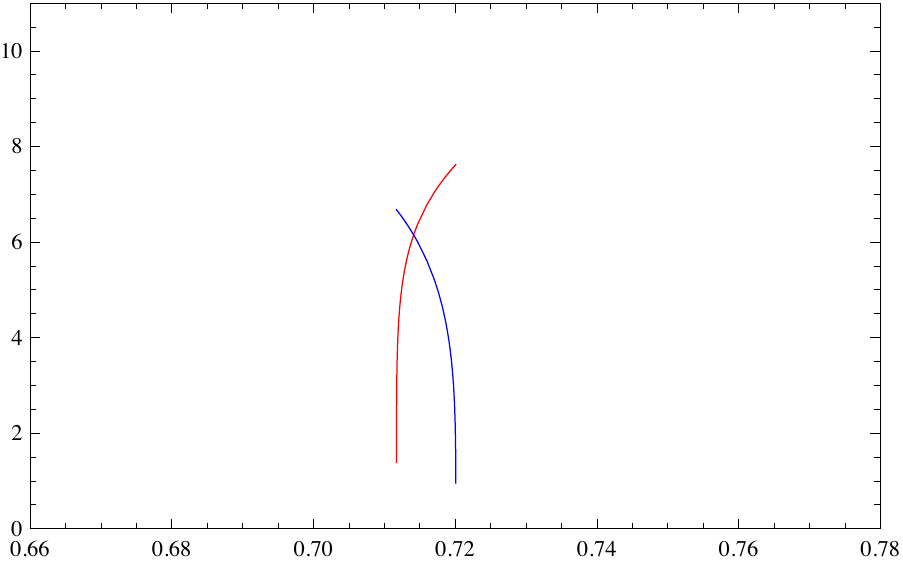}
\put(-100,-10){$T/T_\textrm{d}$}
\put(-320,-10){$T/T_\textrm{d}$}
\put(-430,80){$\frac{\langle\mathcal{O}_\textrm{s}\rangle^{1/2}}{T_\textrm{d}}$}
\put(-213,80){$\frac{\langle\mathcal{O}_\textrm{s}\rangle^{1/2}}{T_\textrm{d}}$}
\put(-213,50){$\frac{\langle\mathcal{O}_\textrm{d}\rangle^{1/3}}{T_\textrm{d}}$}
\put(-430,50){$\frac{\langle\mathcal{O}_\textrm{d}\rangle^{1/3}}{T_\textrm{d}}$}
\caption{Left figure is for the s+d coexistent solution of $\eta=-1/10$ and right figure is for that of $\eta=0$. The red curve is for $\langle\mathcal{O}_\textrm{d}\rangle$ and the blue curve is for $\langle\mathcal{O}_\textrm{s}\rangle$.}
\label{fig2}
\end{figure}
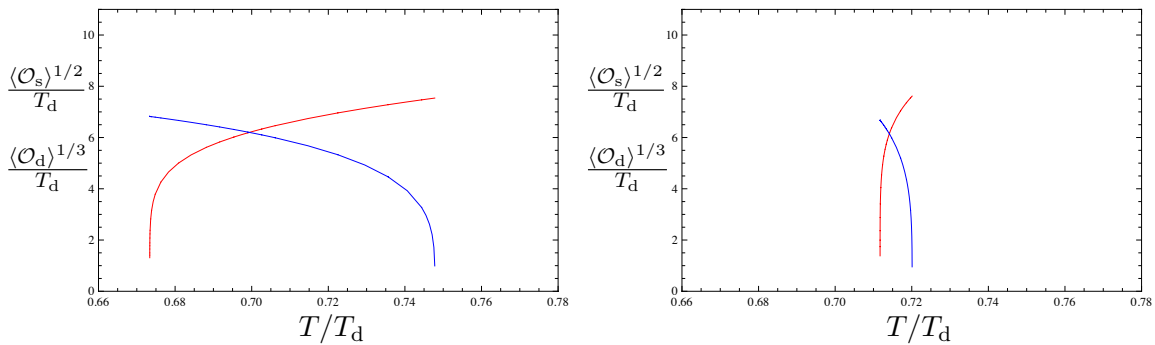
Figure 3 is for the solution of $\eta=1/10$. Unlike figure 2, lowering the temperature, $\langle\mathcal{O}_\textrm{s}\rangle$ becomes small and $\langle\mathcal{O}_\textrm{d}\rangle$ becomes large. From these figures, one can see that the properties of the s+d coexistent solutions depend crucially on the value of $\eta$.

\begin{figure}[t]
\centering
\includegraphics[scale=0.7]{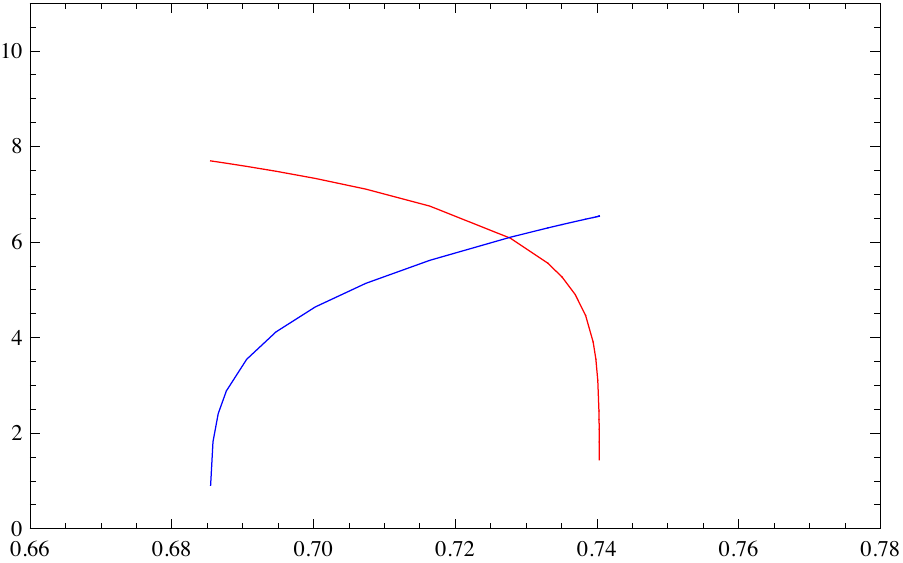} 
\put(-100,-10){$T/T_\textrm{d}$}
\put(-213,80){$\frac{\langle\mathcal{O}_\textrm{s}\rangle^{1/2}}{T_\textrm{d}}$}
\put(-213,50){$\frac{\langle\mathcal{O}_\textrm{d}\rangle^{1/3}}{T_\textrm{d}}$}
\caption{Plot of the s+d coexistent solution of $\eta=1/10$. }
\label{fig3}
\end{figure}   
%%%%%%%%%%%%%%%

\subsection{Free energy density}
In order to see which solution is favored, we compare the free energy densities of the solutions. In holographic superconductor, the free energy corresponds to temperature times the on-shell Euclidean action by assuming the GKP-W relation \cite{Gubser:1998bc, Witten:1998qj}. Thus, we will calculate the on-shell Euclidean action by substituting the solutions.

Usually, the Euclidean action includes the Gibbons-Hawking term \cite{Hawking} and a counter term. They are needed for a well-defined variational principle and dealing with divergences. In our calculation, the contribution of them is same for each solution since we consider the probe limit and (2.19). Therefore, in order to compare the magnitude of the on-shell Euclidean action, it is sufficient to consider the Euclidean (2.4) only. Substituting (2.12), the Euclidean (2.4) is
\begin{align}
S=&-\int dt dx dy\int dz \sqrt{g}\Big[g^{zz}g^{tt}\frac{\phi'^2}{2}+2\psi^2-g^{zz}\psi'^2+g^{tt}e_\textrm{s}^2\phi^2\psi^2+g^{tt}g^{xx}g^{yy}\frac{e_\textrm{d}^2\phi^2\varphi^2}{2z^4}\notag\\
&\;\;\;\;\;\;\;\;\;\;\;\;\;\;\;\;\;\;\;\;\;\;\;\;\;\;\;\;\;\;\;\;\;-g^{zz}g^{xx}g^{yy}\frac{\varphi'^2}{2z^4}-g^{zz}g^{xx}g^{yy}\frac{\varphi^2}{z^6}+(g^{xx}g^{yy})^2\frac{f}{z^8}\varphi^2-\eta g^{xx}g^{yy}\frac{\psi^2\varphi^2}{2z^4}\Big]\notag\\
=&-\int dt dx dy\int dz\Big[\phi'^2/2+2\frac{\psi^2}{z^4}-\frac{f\psi'^2}{z^2}+\frac{e_\textrm{s}^2\phi^2\psi^2}{z^2f}+\frac{e_\textrm{d}^2\phi^2\varphi^2}{2z^2f}-\frac{f\varphi'^2}{2z^2}-\eta\frac{\psi^2\varphi^2}{2z^4}\Big],
\end{align}
where we use $t$ as the imaginary time.
By defining $\beta=\int dt$ and $V_2=\int dx dy$, the on-shell Euclidean action $S_{\textrm{on-shell}}$ is written by
\begin{equation}
\frac{S_{\textrm{onshell}}}{\beta V_2}=-\frac{\mu\rho}{2}+\int\frac{e_\textrm{s}^2\phi^2\psi^2}{z^2f}dz+\int\frac{e_\textrm{d}^2\phi^2\varphi^2}{2z^2f}dz-\int\eta\frac{\psi^2\varphi^2}{2z^4}dz
\end{equation}    
where we have used a partial integration and the equations of motion. (3.2) corresponds to the free energy density because a period of the imaginary time can be interpreted as the thermodynamic $\beta$.

By using the formula (3.2), we compare the free energy densities of the normal conducting solution $F_\textrm{n}$, the s-wave solution $F_\textrm{s}$, the d-wave solution $F_\textrm{d}$ and the s+d coexistent solution $F_\textrm{s+d}$. Then we find following facts:
\begin{itemize}
\item $F_\textrm{n}$ is larger than $F_\textrm{s}$ and $F_\textrm{d}$. Hence, the normal conducting phase is favored at $T>T_\textrm{d}$ only. 
\item $F_\textrm{d}$ is smaller than $F_\textrm{s}$ at high temperature and $F_\textrm{s}$ is smaller than $F_\textrm{d}$ at low temperature. Near $T/T_\textrm{d}=0.716$, the magnitude relation of $F_\textrm{s}$ and $F_\textrm{d}$ is reversed and the s+d coexistence solution exists.
\item $F_\textrm{s+d}$ of $\eta=0$ is smaller than $F_\textrm{s}$ and $F_\textrm{d}$. However, $F_\textrm{s+d}$ of $\eta=1/10$ is larger than $F_\textrm{s}$ and $F_\textrm{d}$. Therefore, the s+d coexistence phase of $\eta=0$ is favored, but that of $\eta=1/10$ is not favored. 
\end{itemize}

Figure 4 (left) is for the free energy densities of $\eta=0$ and figure 4 (right) is for that of $\eta=1/10$. The blue line is for $F_\textrm{s}$, the red line is for $F_\textrm{d}$, the orange line is for $F_\textrm{s+d}$ of $\eta=0$ and the green line is for that of $\eta=1/10$. From these figures, we conclude that the s+d coexistence phase of $\eta=0$ is favored, but that of $\eta=1/10$ is not favored. The relation among the free energy densities for each solution is explained again in figure 6.
\begin{figure}[t]
\centering
\includegraphics[scale=0.55]{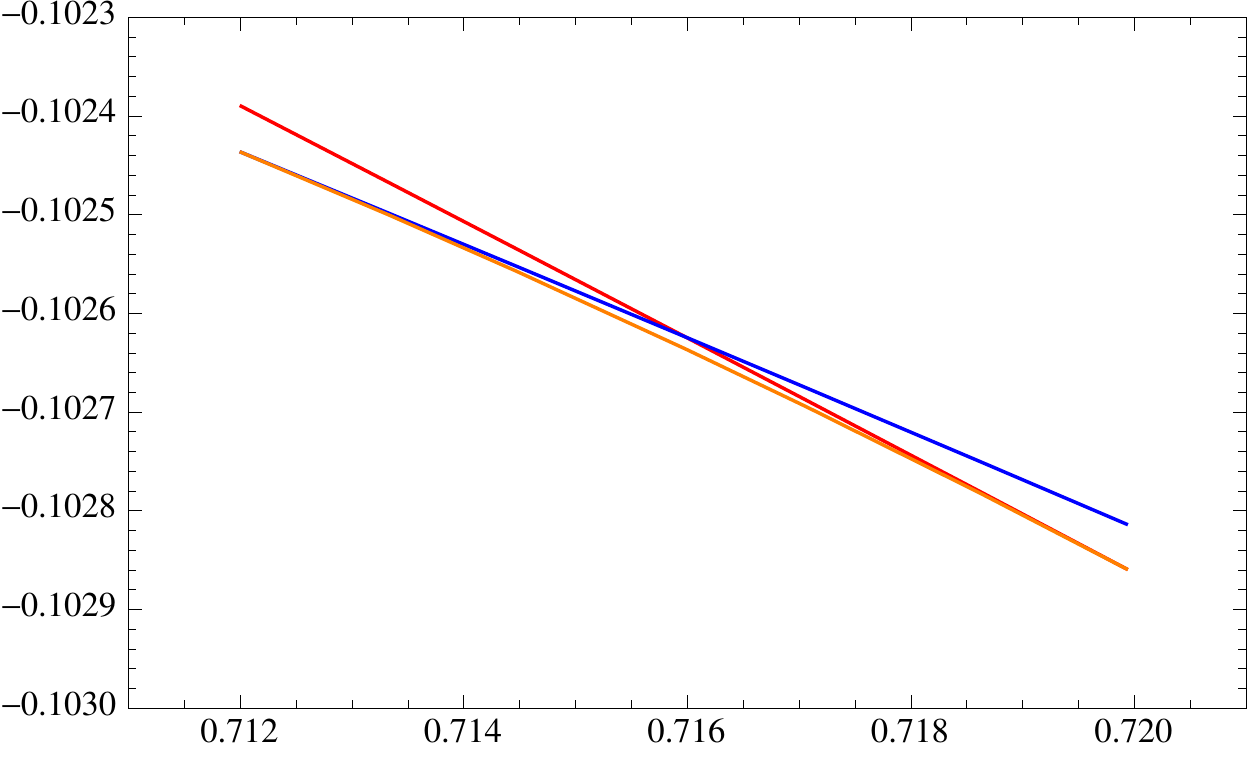} \hspace{8mm}
\includegraphics[scale=0.55]{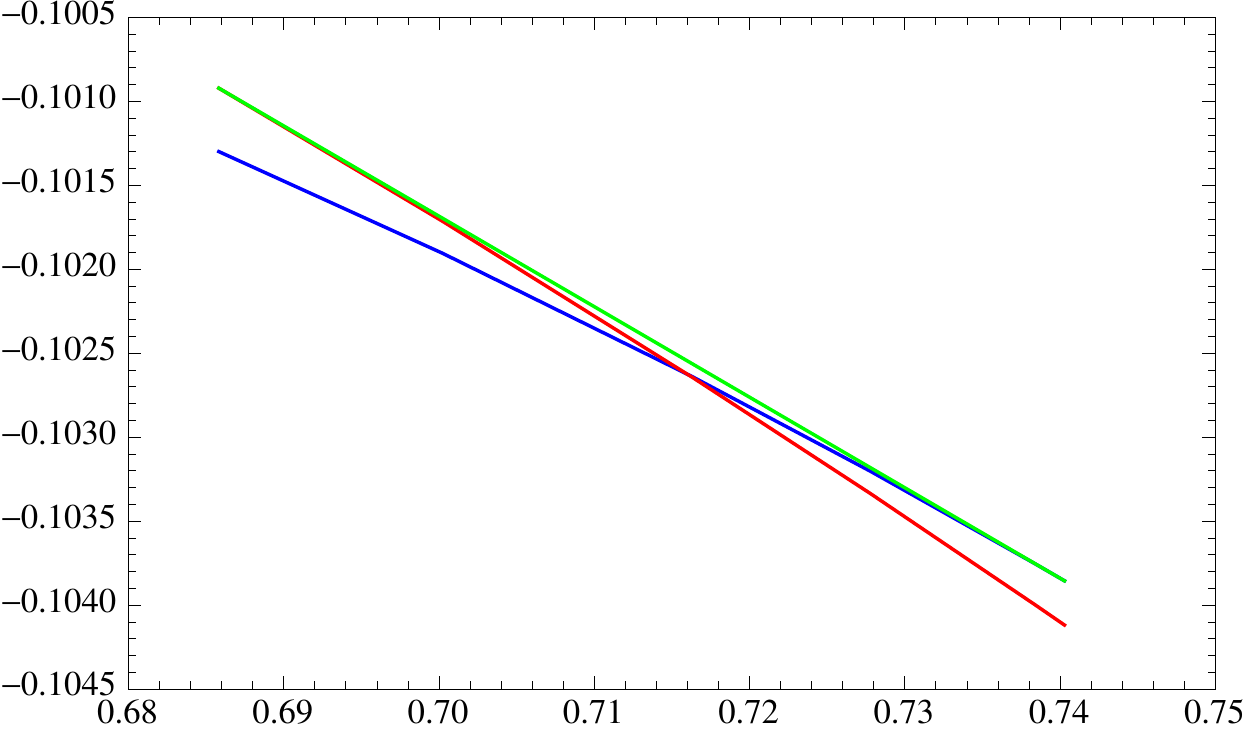}
\put(-100,-10){$T/T_\textrm{d}$}
\put(-320,-10){$T/T_\textrm{d}$}
\put(-458,60){$\frac{S_\textrm{onshell}}{\beta V_2}$}
\put(-228,60){$\frac{S_\textrm{onshell}}{\beta V_2}$}
\caption{Left figure is for the free energy densities of $\eta=0$ and right figure is for that of $\eta=1/10$. The blue line is for $F_\textrm{s}$, the red line is for $F_\textrm{d}$, the orange line is for $F_\textrm{s+d}$ of $\eta=0$ and the green line is for that of $\eta=1/10$.}
\label{fig4}
\end{figure}
\section{Phase diagram}

In this section, we study a phase diagram of our model by using the property of the free energy density.

Figure 5 is for the order parameters in the favored phases at $T$. Figure 5 (left) is for the phase of $\eta=0$. Lowering the temperature, the phase is changed in the following order: the normal conducting phase, the d-wave phase, the s+d coexistence phase and the s-wave phase. In these phase transitions, the order parameters change continuously. This figure corresponds to the figure 5 of \cite{Basu:2010fa} because the direct coupling $\eta$ is zero and there are four phases.

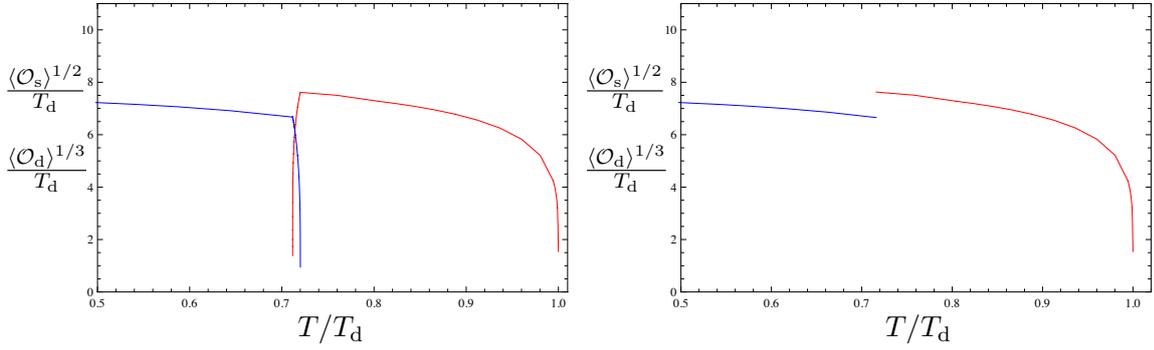
\begin{figure}[t]
\centering
\includegraphics[scale=0.7]{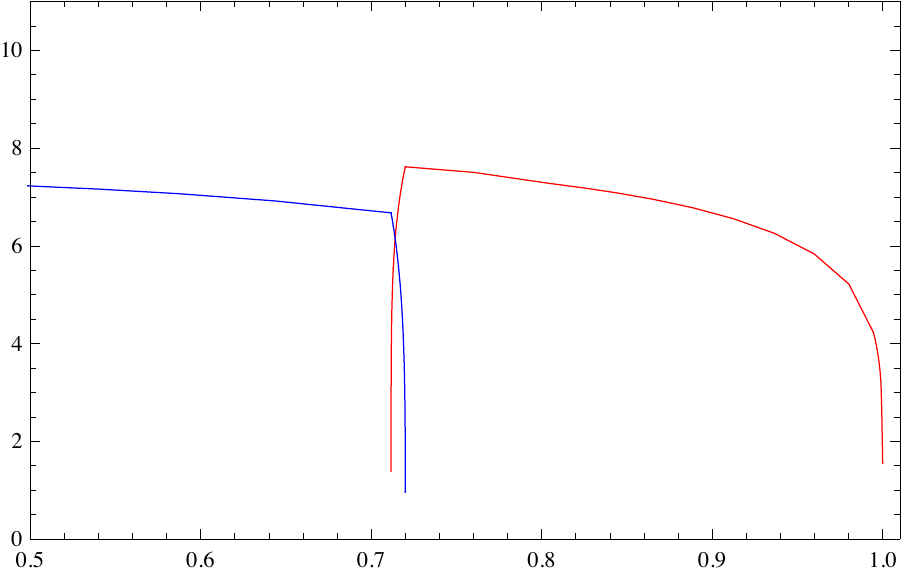} \hspace{10mm}
\includegraphics[scale=0.7]{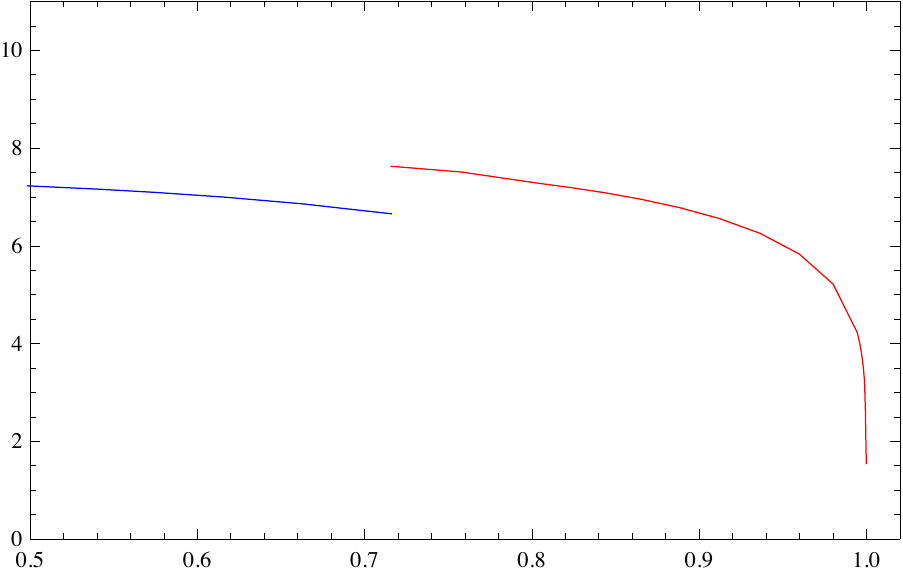}
\put(-100,-10){$T/T_\textrm{d}$}
\put(-320,-10){$T/T_\textrm{d}$}
\put(-430,80){$\frac{\langle\mathcal{O}_\textrm{s}\rangle^{1/2}}{T_\textrm{d}}$}
\put(-213,80){$\frac{\langle\mathcal{O}_\textrm{s}\rangle^{1/2}}{T_\textrm{d}}$}
\put(-213,50){$\frac{\langle\mathcal{O}_\textrm{d}\rangle^{1/3}}{T_\textrm{d}}$}
\put(-430,50){$\frac{\langle\mathcal{O}_\textrm{d}\rangle^{1/3}}{T_\textrm{d}}$}
\caption{Left figure is the plot of the order parameters for the favored phase of $\eta=0$ and right figure is that of $\eta=1/10$. The s+d coexistent phase is found for $\eta=0$, but that for $\eta=1/10$ is not favored.}
\label{fig5}
\end{figure}

Figure 5 (right) is for the phase of $\eta=1/10$. Lowering the temperature, the phase changes in the following order: the normal conducting phase, the d-wave phase and the s-wave phase. In the latter phase transition, the order parameters change discontinuously since the s+d coexistence phase (figure 3) is not favored.

The physics of figure 5 can be explained as follows. Since a large $e_\textrm{d}$ makes the effective mass squared of the d-wave small, the d-wave phase condenses first. Lowering the temperature, the effect of $m_\textrm{s}^2$ becomes important and the s-wave phase appears. There is a case that the s+d coexistence phase exists to connect the two phases continuously. 

In the Lagrangian, $\eta\psi^2\varphi^2/2$ is the term corresponding to the potential energy. If $\eta$ is small enough, the s+d coexistence solution is advantageous for energy. Thus, the range of $T$ in which the s+d coexistence solution of $\eta=-1/10$ exists is larger than that of $\eta=0$. If $\eta$ is large enough, the s+d coexistence phase cannot exist, and one example shown in figure 5 is for $\eta=1/10$.  

Figure 6 is for the free energy of each solution at $T$. This figure is a rough sketch and the scale is not correct. The blue line is for $F_\textrm{s}$, the red line is for $F_\textrm{d}$, the purple line is for $F_\textrm{s+d}$ of $\eta=-1/10$, the orange line is for that of $\eta=0$ and the green line is for that of $\eta=1/10$. The  solution which is favored at a given $T$ corresponds to  the lowest line since the free energy is the smallest. The s+d coexistence solution of small $\eta$ (purple) exists at the lower left. Increasing $\eta$, the solution (orange) moves to the upper right. If $\eta$ is large enough, the solution (green) has a free energy larger than that of the single solutions, and the s+d coexistence phase cannot exist.
\begin{figure}[t]
\centering
\includegraphics[scale=0.3]{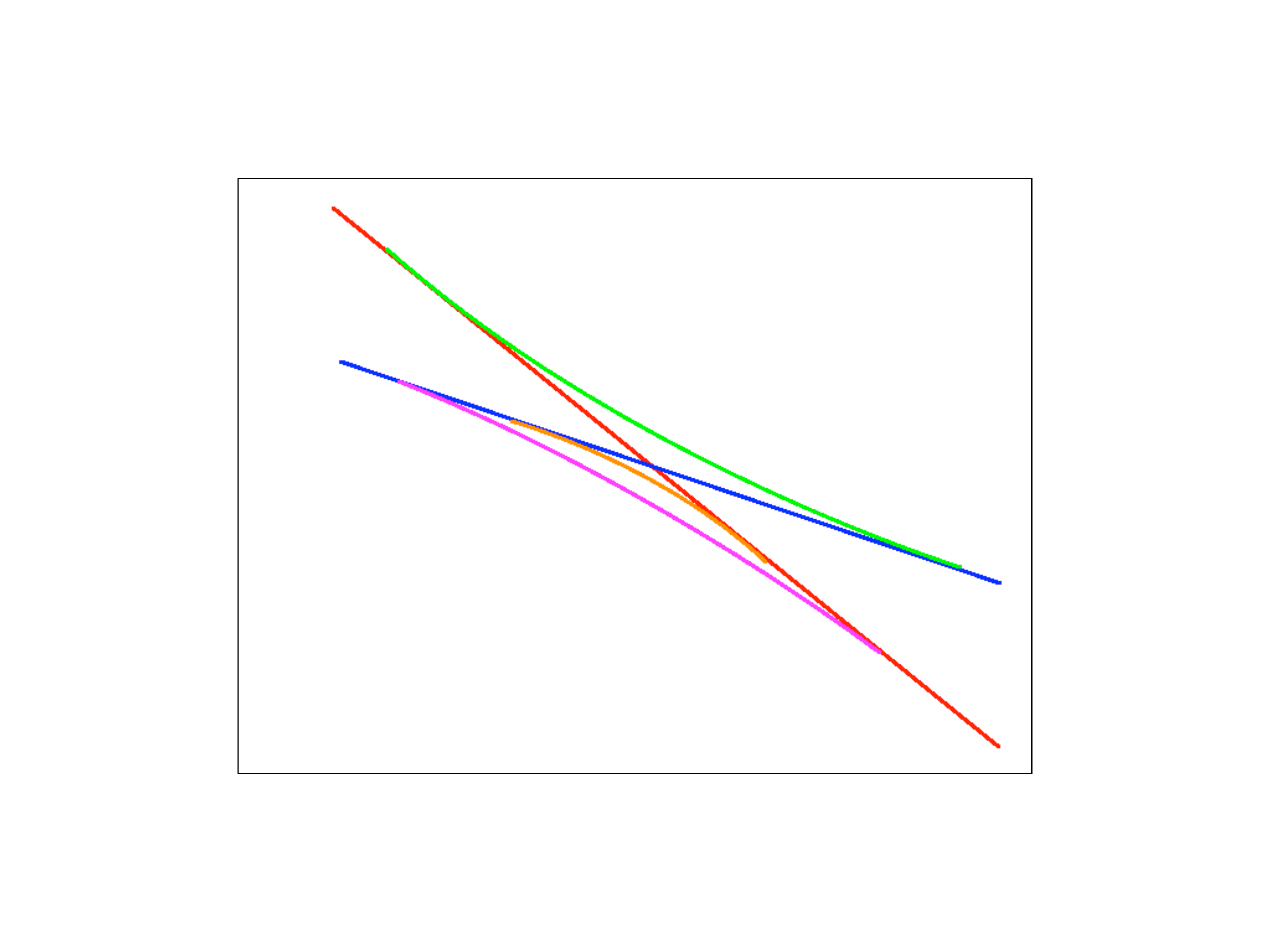} 
\put(-230,70){$\frac{S_\textrm{onshell}}{\beta V_2}$}
\put(-100,-15){$T/T_\textrm{d}$}
\caption{A rough sketch of the free energy of each solution. The blue line is for $F_\textrm{s}$, the red line is for $F_\textrm{d}$, the purple line is for $F_\textrm{s+d}$ of $\eta=-1/10$, the orange line is for that of $\eta=0$ and the green line is for that of $\eta=1/10$.}
\label{fig6}
\end{figure}

By summarizing the above results, we can draw a phase diagram. Figure 7 is the $\eta$-$T$ phase diagram of favored states. The green line is for $T_\textrm{d}$. The red curve is for the temperature at which the s+d coexistence phase starts to appear as lowering $T$, and the blue curve is for that at which the s+d coexistence phase ends. In these curves, the black dots are for our numerical results, and we simply connect them by lines. The purple line means a first order phase transition between the single solutions. There are four phases corresponding to four solutions in figure 7, the normal conducting phase, the s-wave single phase, the d-wave single phase and the s+d coexistence phase. When the temperature passes the green, red or blue line, a phase transition at which $\langle\mathcal{O}_\textrm{s}\rangle$ and $\langle\mathcal{O}_\textrm{d}\rangle$ change continuously occurs. Otherwise, when the temperature passes the purple line, the phase transition at which they change discontinuously occurs. 

\begin{figure}[t]
\centering
\includegraphics[scale=0.7]{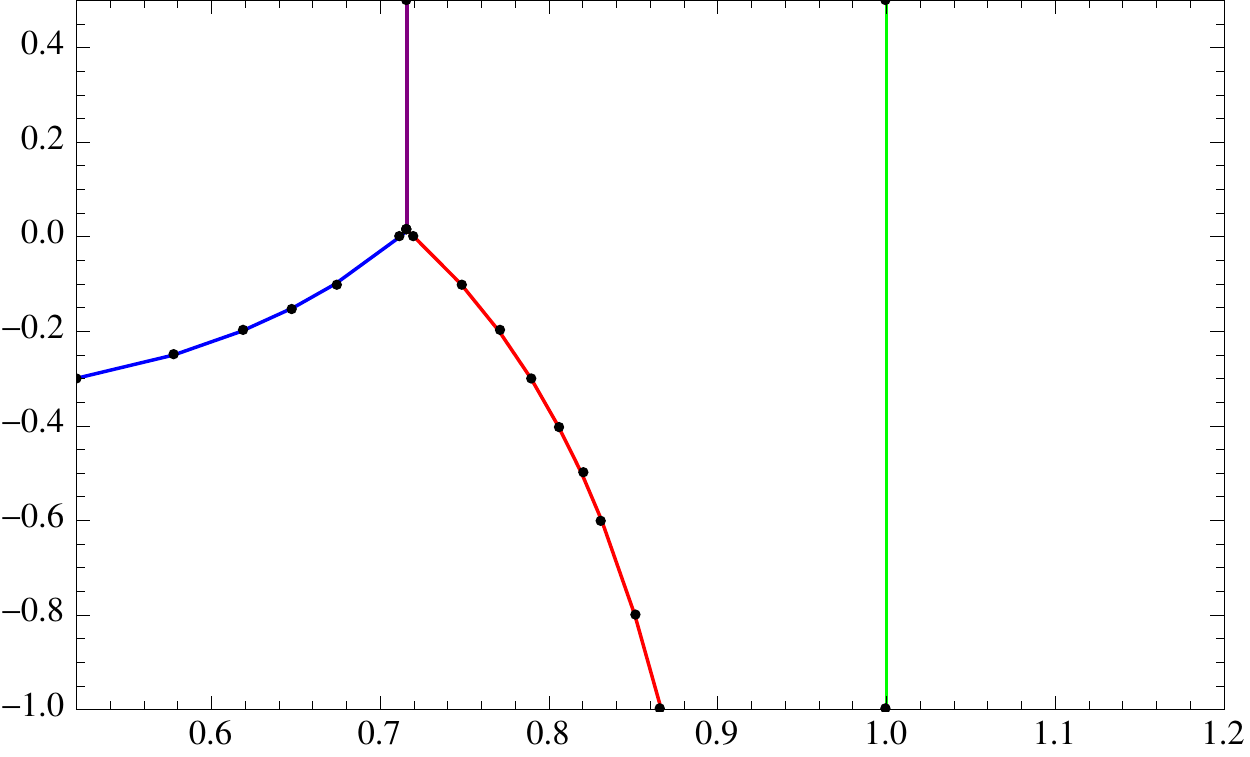} 
\put(-265,85){$\eta$}
\put(-120,-15){$T/T_\textrm{d}$}
\put(-55,80){normal}
\put(-130,110){d-wave}
\put(-225,110){s-wave}
\put(-190,50){s+d}
\caption{$\eta$-$T$ phase diagram of favored states. $\eta$ is the direct coupling between the s-wave and the d-wave. The green line is for $T_\textrm{d}$. The  red curve is for the temperature at which the s+d coexistence phase starts to appear as lowering $T$, and the blue curve is for that at which the s+d coexistence phase ends. The purple line means a first order phase transition between the single solutions.}
\label{fig7}
\end{figure}

\begin{figure}[t]
\centering
\includegraphics[scale=0.5]{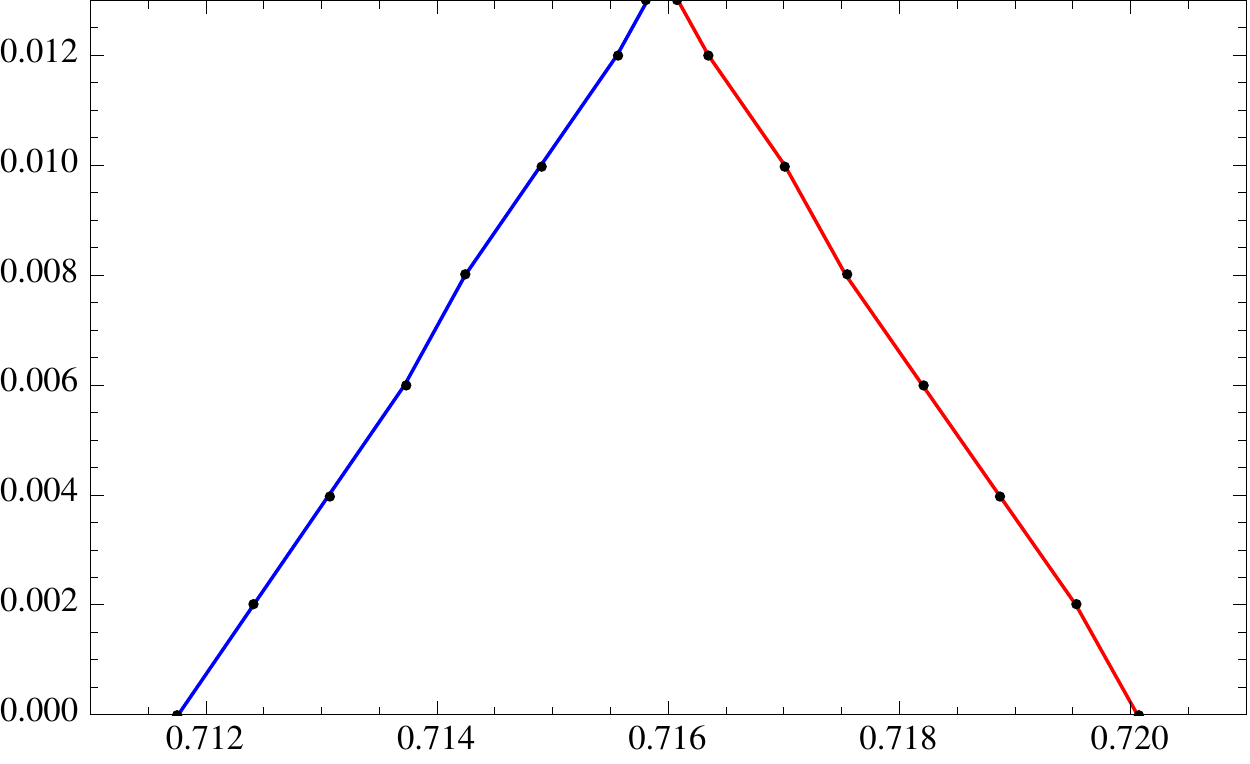} 
\put(-190,70){$\eta$}
\put(-100,-15){$T/T_\textrm{d}$}
\put(-45,70){d-wave}
\put(-160,70){s-wave}
\put(-95,40){s+d}
\caption{$\eta$-$T$ phase diagram near $\eta=0$.}
\label{fig8}
\end{figure}
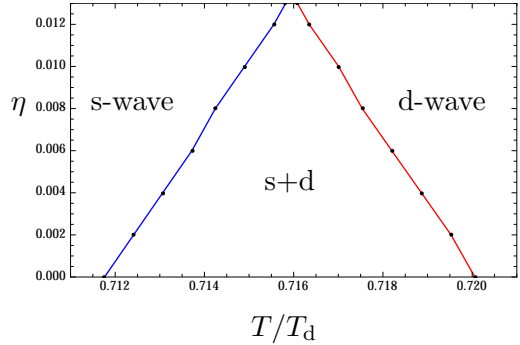

Figure 8 is the phase diagram near $\eta=0$. From this figure, one can see that the red and blue line intersect at $T/T_\textrm{d}\cong0.716$. This value of the temperature is same as the purple line. Therefore, we can conclude that there is a triple point in this phase diagram. 

Finally, we discuss the phase diagram with other parameters. Figure 9 and figure 10 are the phase diagrams with the parameters as\footnote{We choose $m^2_\textrm{d}L^2=4$ and $e_\textrm{d}=2.9$ for simplicity of boundary behavior and coexistence.}
\begin{align}
&m^2_\textrm{s}L^2=-2,\;\;\;\;\;m^2_\textrm{d}L^2=4,\\
&e_\textrm{s}=1,\;\;\;\;\;\;\;\;\;\;\;\;\;\;e_\textrm{d}=2.9.
\end{align}
Topology of figure 9 is same as that of figure 7, therefore, we expect that the phase diagrams have same features like a triple point even if $m^2_\textrm{d}$ is large.

\begin{figure}[t]
\centering
\includegraphics[scale=0.75]{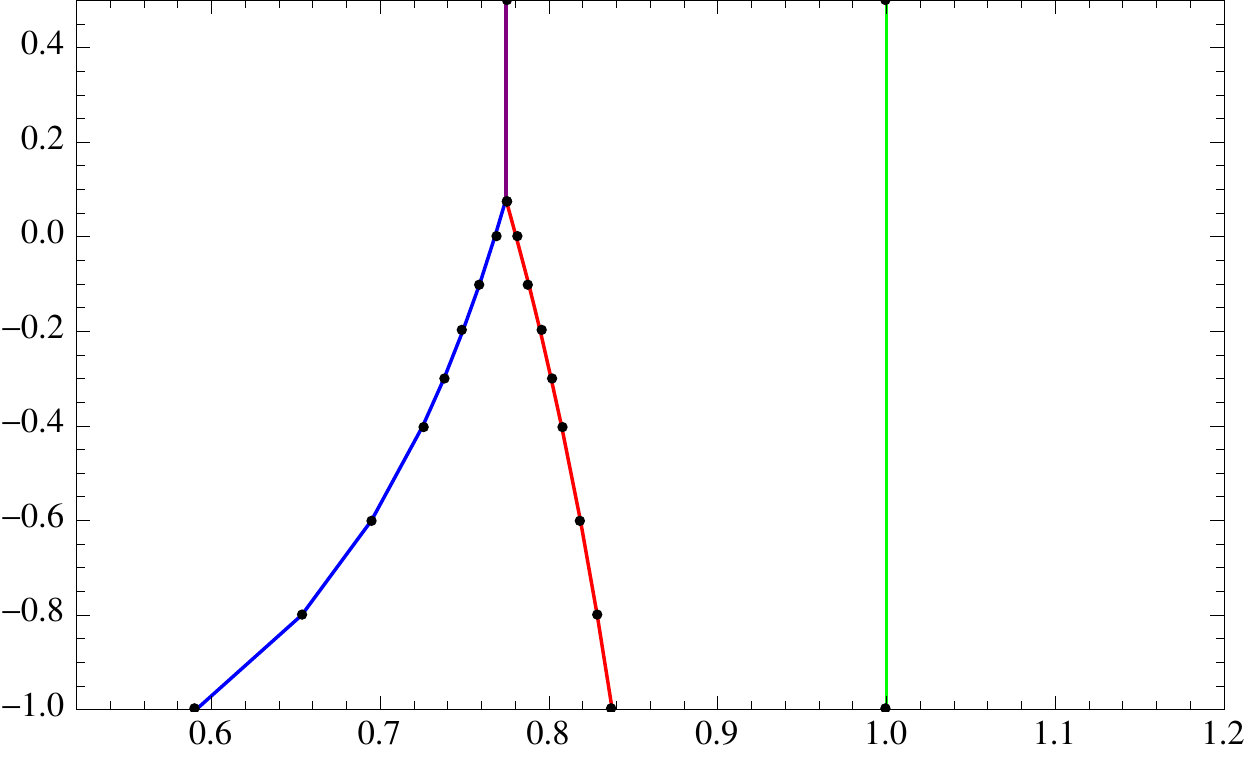} 
\put(-285,90){$\eta$}
\put(-125,-15){$T/T_\textrm{d}$}
\put(-55,90){normal}
\put(-130,90){d-wave}
\put(-225,90){s-wave}
\put(-175,50){s+d}
\caption{$\eta$-$T$ phase diagram with the parameters (4.1) and (4.2).}
\label{fig9}
\end{figure}

\begin{figure}[t]
\centering
\includegraphics[scale=0.5]{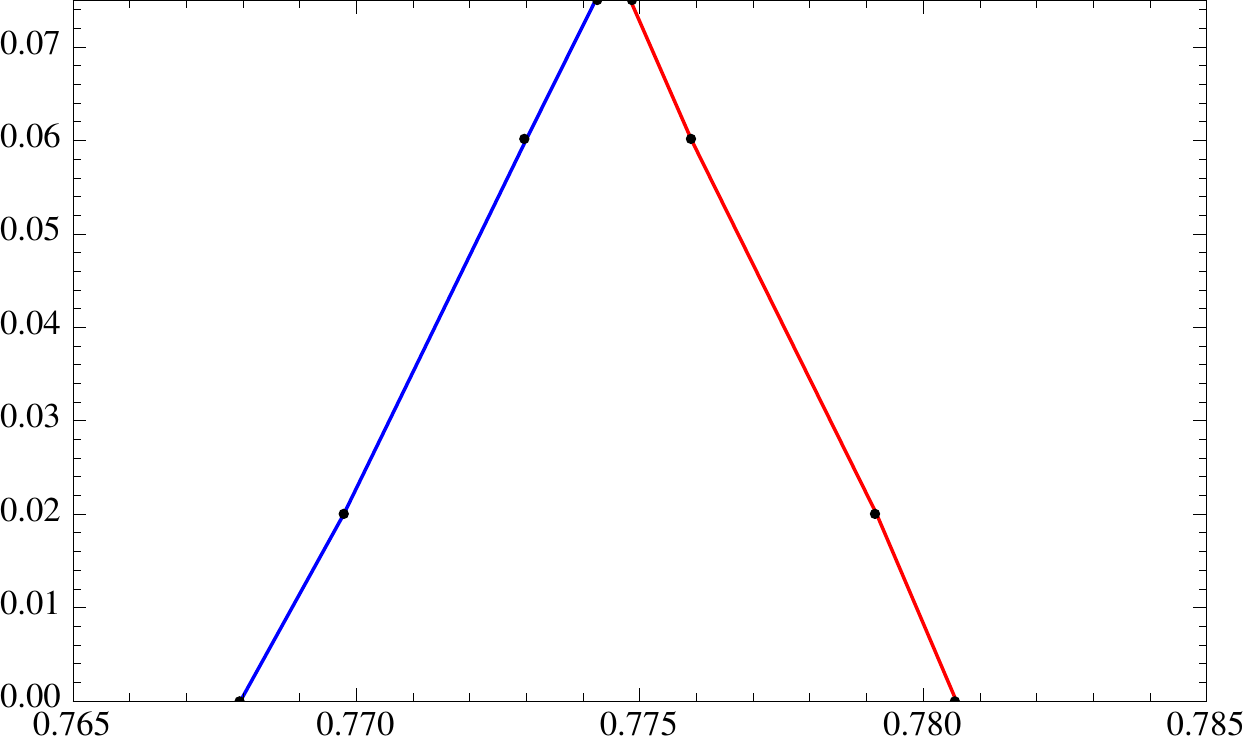} 
\put(-190,70){$\eta$}
\put(-100,-15){$T/T_\textrm{d}$}
\put(-50,70){d-wave}
\put(-160,70){s-wave}
\put(-100,40){s+d}
\caption{$\eta$-$T$ phase diagram near $\eta=0$ with the parameters (4.1) and (4.2).}
\label{fig10}
\end{figure}
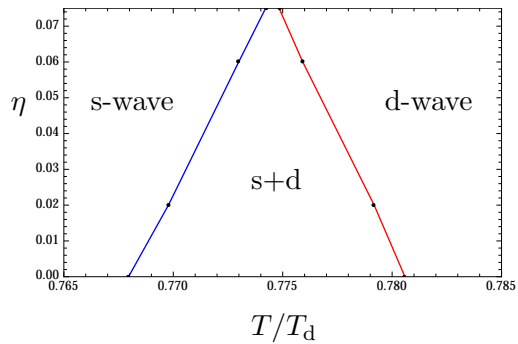

%%%%%%%%%%%%%%%%
\section{Summary and discussion}

In this paper, we have calculated the solutions of the holographic superconductor model with an s-wave and a d-wave (2.4) in the probe limit. We consider the direct coupling only as this model does not accommodate the Josephson coupling which exist in the two-scalar model \cite{Wen:2013ufa}. We have found that the model has a rich phase structure: 
\begin{itemize}
\item continuous or discontinuous phase transition of the order parameters 
\item coexistence of the s-wave and the d-wave
\item an order competition
\item a triple point
\end{itemize}

Under a certain ansatz for the fields, our holographic superconductor model reduces to a two-scalar model. However, if we consider some other coupling between the s-wave and d-wave or some other ansatz for the fields, the phase diagram perhaps changes.  

There are some future issues: 

\begin{itemize}
\item Calculating the model with a back reaction or some other coupling. For the calculation, it is important to study a higher spin theory on curved spacetimes.
\item Interpreting $e_\textrm{s}$ and $e_\textrm{d}$ as physical observables in superconductors.  In our model, we have chosen different values for $e_\textrm{s}$ and $e_\textrm{d}$, while it is not clear whether the chosen parameters can be physically realized.
\item Estimating an error of the numerical calculations. Although a possible change by the detailed numerical calculations does not change the phase diagram because $F_\textrm{s+d}$ of $\eta=1/10$ is larger than $F_\textrm{d}$, it would be a little strange that there is a range of the temperature that $F_\textrm{s+d}$ of $\eta=1/10$  is smaller than $F_\textrm{s}$. Therefore, it may be meaningful to check whether the range is caused by an error of the numerical calculation or not. 
\end{itemize}

Other future work includes, for example, a calculation of a conductivity, a detailed calculation of the blue and red curve of figure 6, and an analysis of the instability of the model about $\eta$. Furthermore, we expect that a holographic model with bosonic and fermionic degrees of freedom and a Yukawa coupling \cite{Liu:2013yaa} has the property similar to our model.  

%%%%%%%%%%%%%%%%

\acknowledgments

I would like to thank K. Hashimoto for helpful discussion and careful reading of the manuscript. I would also like to thank Y. Hosotani, T. Onogi, A. Tanaka and S. Yamaguchi for useful comments.

%%%%%%%%%%%%%%%%%%%%%%%%%%%%%%%%%%%%%%%%%

\end{document}